\documentclass[12pt]{article}
\newcommand{\Comment}[1]{{}}
\usepackage[textwidth = 430 pt, textheight = 630 pt]{geometry}
\usepackage{amssymb,euscript,amsmath,amsfonts}

\usepackage{color}
\definecolor{MyDarkBlue}{rgb}{0.15,0.15,0.45}
\usepackage[linktocpage=true]{hyperref}
\hypersetup{
colorlinks=true,
citecolor=MyDarkBlue,
linkcolor=MyDarkBlue,
urlcolor=MyDarkBlue,
pdfauthor={Neil Lambert},
pdftitle={Heterotic M2's},
pdfsubject={hep-th}
}

\usepackage[numbers,sort&compress]{natbib}
\usepackage{hypernat}
\usepackage{url}

\newcommand\ignore[1]{}
\def\one{{\,\hbox{1\kern-.8mm l}}}

\newcommand{\Cset}{{\,\,{{{^{_{\pmb{\mid}}}}\kern-.45em{\mathrm C}}}}}

\newcommand{\be}{\begin{equation}}
\newcommand{\ee}{\end{equation}}
\newcommand{\bea}{\begin{eqnarray}}
\newcommand{\eea}{\end{eqnarray}}

\begin{document}

\renewcommand{\thefootnote}{\fnsymbol{footnote}}

\rightline{KCL-MTH-15-05}

   \vspace{1.8truecm}

 \centerline{\LARGE \bf {\sc Heterotic M2-Branes}}

\centerline{\LARGE \bf {\sc  }} \vspace{2truecm} \thispagestyle{empty} \centerline{
    {\large {\bf {\sc Neil~Lambert${}^{\,}$}}}\footnote{E-mail address: \href{mailto:neil.lambert@kcl.ac.uk}{\tt neil.lambert@kcl.ac.uk}}{,}
                                                          }

\vspace{1cm}
\centerline{{\it Department of Mathematics}}
\centerline{{\it King's College London, WC2R 2LS, UK}}

\vspace{1.0truecm}

 
\thispagestyle{empty}

\centerline{\sc Abstract}
\vspace{0.4truecm}
\begin{center}
\begin{minipage}[c]{360pt}{
    \noindent    We construct the action for $N$ M2-branes on $S^1/\mathbb{Z}_2$. The resulting theory has a gauge anomaly but this can be cancelled if the two fixed point planes each support 8 chiral Fermions in the fundamental of $U(N)$. Taking the low energy limit leads to the worldsheet theory of $N$ free heterotic strings whose quantization induces an $E_8$ spacetime gauge symmetry on each fixed point plane. Thus this paper presents a non-abelian worldvolume analogue of the classic Ho$\check{{\rm r}}$ava-Witten analysis.}

\end{minipage}
\end{center}

\newpage 
 
\renewcommand{\thefootnote}{\arabic{footnote}}
\setcounter{footnote}{0}

\section{Introduction}

M-theory is still a somewhat mysterious theory with no satisfactory microscopic description. Formally it can be thought of as the strong coupling limit of type IIA string theory. As such fundamental strings lift to wrapped M2-branes. Putting this the other way M-theory on $S^1$ gives type IIA string theory and wrapped M2-branes become fundamental strings. A variation of this is the striking result that M-theory on an interval, viewed as an obifold $S^1/{\mathbb Z}_2$, gives the $E_8\times E_8$ heterotic string \cite{Horava:1995qa,Horava:1996ma}. From the M-theory point of view the appearance of a dynamical $E_8\times E_8$ spacetime  gauge bundle  is somewhat magical. Its existence is inferred from the need to cancel ten-dimensional spacetime gauge anomalies, along with the fact that there are two  fixed points planes so that the gauge group must factorize into two equal components. This selects the $E_8\times E_8$ heterotic string over the $Spin(32)/{\mathbb Z}_2$ one.
 
The the aim of this letter is to construct the non-abelian action for multiple heterotic strings from an $S^1/{\mathbb Z}_2$ orbifold of $N$ M2-branes, along  with additional twisted sector states. Reducing to two dimensions this leads to a theory of $N$ free heterotic strings. In a sense this is a non-abelian  worldvolume analogue of the Ho$\check{{\rm r}}$ava-Witten analysis \cite{Horava:1995qa,Horava:1996ma}  which was largely based on spacetime anomalies, although   \cite{Horava:1995qa} also gave an argument for the existence of a chiral $c=16$ twisted sector in the abelian M2-brane theory using worldvolume gravitational anomalies. Furthermore in \cite{Rey:1997hj,Kabat:1997za} the existence of  twisted sector fermions was identified using anomalies in a Matrix construction of the heterotic string. In our analysis we must again appeal to the logic that consistency of  M-theory on an orbifold implies the existence of new  states that are localized at the fixed points, without a true microscopic understanding. Nevertheless we hope that this analysis helps to shed more light on the origin of $E_8$ structure from an M-theory perspective. In particular it has the arguably  less exotic aim that we need only look for additional chiral Fermion modes at each fixed point, with no apparent non-abelian structure. The $E_8$ spacetime symmetry on the fixed point planes then arises from quantization of the worldsheet fermions in a well-known way \cite{Gross:1985fr,Gross:1985rr}.

The extra fermionic degrees of freedom that we require should only arise in the case of an orbifold $S^1/{\mathbb Z}_2$: simply putting an M2-brane on an interval, for example if the M2-branes are suspended between two M5-branes, is not expected to lead spacetime $E_8$ gauge symmetry. Otherwise it would show also up in the dynamics of the $(2,0)$ theory and hence on D4-branes. Thus our analysis is complimentary to the boundary conditions  considered in \cite{Berman:2009xd,Chu:2009ms}. On the other hand these fermionic degrees of freedom will arise in the case of the $(1,0)$ E-string theories that have recently been studied  in \cite{Kim:2014dza,Haghighat:2014pva}. M2-brane anomalies have also recently featured in \cite{witten:strings}. And we hope our analysis will be useful for these theories.

The rest of this letter is organised as follows. In section 2 we will review the worldvolume theory of $N$ M2-branes and construct an orbifold of it under a worldvolume parity transformation, resulting in a two-dimensional Kaluza-Klein (KK) theory with a single $U(N)$ gauge group. In section 3 we will argue that the this two-dimensional theory
has a gauge anomaly which can be cancelled if each of the two fixed-point planes  supports 8 chiral modes in the fundamental $U(N)$. In section 4 we will take the low energy limit of the anomaly-free theory and show that it reduces to $N$ copies of worldvolume theory the heterotic string  with $SO(16)\times SO(16)$ symmetry, leading to spacetime $E_8\times E_8$ gauge symmetry in ten dimensions. In section 5 we state our conclusions.
 
 \section{Orbifolding M2's}
 
 Let us recall the action for $N$ M2-branes \cite{Aharony:2008ug} in an ${\mathbb C^4}/{\mathbb Z}_k$ transverse space obtained as a $U(N)\times U(N)$ Chern-Simons Matter theory:
 \begin{align}
 S = -{\rm tr} \int d^3 x&\Big\{\  D_m Z^A D^m Z_A +\frac{8\pi^2}{3k^2}\Upsilon^{CD}_B\Upsilon_{CD}^B\nonumber\\
 &+\frac{k}{4\pi} \left(A^L_m\partial_n A^L_p -\frac{2i}{3}A^L_mA^L_nA^L_p\right) - 
 \frac{k}{4\pi} \left(A^R_m\partial_n A^R_p -\frac{2i}{3}A^R_mA^R_nA^R_p\right)\nonumber\\
 &+i \bar\psi^A\gamma^m D_m \psi_A+\frac{2i\pi}{k} \bar\psi^A[\psi_A,Z^B;Z_B]-\frac{4i\pi}{k} \bar\psi^A[\psi_B,Z^B;Z_A]\nonumber\\
  &-\frac{i\pi}{ k}\varepsilon_{ABCD}\psi^A[Z^C,Z^D;\psi^B]+ \frac{i\pi}{ k}\varepsilon^{ABCD}\psi_A[Z_C,Z_D;\psi_B]
  \Big\}\ ,
 \end{align}
where we have used the construction of \cite{Bagger:2008se} with
\begin{equation}
\Upsilon^{CD}_B =[Z^C,Z^D;Z_B] - \frac12 \delta^C_B[Z^E,Z^D;Z_E] + \frac12 \delta^D_B[Z^E,Z^C;Z_E] \ ,
\end{equation}
\begin{equation}
[Z^A,Z^B;Z_C] = Z^AZ_CZ^B - Z^BZ_CZ^A\ .
\end{equation}
and, {\it e.g.}
\begin{equation}
D_m Z^A = \partial_m Z^A -i A^L_mZ^A + iZ^AA^R_m\ .
\end{equation}
In these expressions the matter fields are $N\times N$ complex matrices and the gauge fields are hermitian $N\times N$ matrices. Furthermore  hermitian conjugation acts by raising/lowering the R-symmetry index $A=1,2,3,4$. We use a convention where $m=0,1,2$ and $\gamma_m$ are $2\times 2$ real matrices.   The same action also describes the original, maximally supersymmetric, model of \cite{Gustavsson:2007vu,Bagger:2007jr} for two M2-branes if the gauge group is taken to be $SU(2)\times SU(2)$.  For a review of these theories see \cite{Bagger:2012jb}.

The action is invariant under  the ${\cal N}=6$ supersymmetry transformations:
\begin{align}
\delta_{\epsilon}Z^A & = i\bar\epsilon^{AB}\psi_B\nonumber\\
\delta_{\epsilon}A^L_m & =  \frac{2\pi}{k}\bar\epsilon^{AB}\gamma_m \psi_AZ_B -  \frac{2\pi}{k}\bar\epsilon_{AB}\gamma_m Z^B \psi^A\nonumber\\
\delta_{\epsilon} A^R_m & =  \frac{2\pi}{k}\bar\epsilon^{AB}\gamma_m Z_B \psi_A- \frac{2\pi}{k}\bar\epsilon_{AB}\gamma_m \psi^AZ^B\nonumber\\
\delta_{\epsilon}\psi_B & = \gamma^m D_m Z^A\epsilon_{AB} + \frac{2\pi}{k}\Upsilon^{CD}_B\epsilon_{CD} \ ,
\end{align}
where $\epsilon^{AB} = \frac{1}{2}\varepsilon^{ABCD}\epsilon_{CD}$.

We need to find a suitable notion of parity on the fields under $x^2\to -x^2$. Since the Chern-Simons terms are parity odd, naively this can be corrected by sending $k\to  -k$. However this is not a symmetry of the theory since the coupling constant is changed. More correctly one thinks of swapping the two $U(N)$ gauge groups. In \cite{Aharony:2008ug} parity was defined as $x^2\to -x^2$ and 
\begin{align}\label{parity1}
Z^A(x^2)&\to (Z^A(-x^2))^\dag\nonumber\\
\psi_A(x^2) &\to \gamma_2(\psi_A(-x^2))^\dag\nonumber\\
A^{L/R}_\mu (x^2)  &\to A^{L/R}_\mu (-x^2)\qquad \mu=0,1\nonumber\\
A^{L/R}_2(x^2) &\to  -A^{R/L}_2(-x^2) \ .
\end{align}
  However if we think of $Z^A = X^A+iX^{A+4}$, with $X^I$ hermitian, then this also involves a reflection in the $x^7,x^8,x^9,x^{10}$ directions, corresponding to an $O(6)$-plane rather than the $O(10)$-plane that we wish to consider. Therefore in this paper we will consider the following action of parity:
\begin{align}
Z^A(x^2)&\to (Z^A(-x^2))^t\nonumber\\
\psi_A(x^2) &\to \gamma_2(\psi_A(-x^2))^t\nonumber\\
A^{L/R}_\mu (x^2)  &\to -(A^{L/R}_\mu (-x^2))^t\qquad \mu=0,1\nonumber\\
A^{L/R}_2(x^2) &\to  (A^{R/L}_2(-x^2))^t\ .
\end{align}
A straightforward calculation then shows that this is indeed a symmetry of the action.
 
If we impose the orbifold, corresponding to states that invariant under $x^2\leftrightarrow -x^2$, then we must restrict to fields that satisfy
\begin{align}\label{oaction}
Z^A(-x^2) & = (Z^A(x^2))^t\nonumber\\
\psi_A(-x^2) & = \gamma_2 (\psi_A(x^2))^t\nonumber\\
A_\mu^{L/R}(-x^2) & = -(A_\mu^{R/L}(x^2))^t\nonumber\\
A_2^{L/R}(-x^2) & = (A_2^{R/L}(x^2))^t \ .
\end{align} 
This breaks half of the supersymmetry as
\begin{align}
\delta_{\epsilon} Z^A(-x^2) & = i\bar\epsilon^{AB}\psi_B(-x^2)\nonumber\\
& = i\bar\epsilon^{AB}\gamma_2(\psi_B(x^2))^t\nonumber\\
& = -i{(\overline  {\gamma_2\epsilon})}^{AB}\psi_B(x^2)\ .
\end{align}
Thus $\delta_{\epsilon}Z^A(-x^2) = (\delta_{\epsilon}Z^A( x^2))^t $ if and  only if 
\begin{equation}
\gamma_2\epsilon^{AB} = -\epsilon^{AB}\ .
\end{equation}
One can then verify that all the other supersymmetries generated by $\epsilon^{AB}_-$ respect the orbifold conditions. 

The orbifold also breaks the $U(N)\times U(N)$ gauge group which acts, for example, on $Z^A$ as $Z^A \to g_LZ^Ag_R^\dag$. One finds that the surviving gauge symmetries satisfy
\begin{equation}
g_R(x^2)=g_L^*(-x^2) \ .
\end{equation}
Thus there is just a single $U(N)$  gauge group, which, from the point of view of the three-dimensional theory, acts non-locally:
\begin{equation}
Z^A(x^2) \to g(x^2)Z^A(x^2)g^t(-x^2)\ .
\end{equation}
Although it acts locally at the fixed points.

We can solve  the orbifold conditions by considering the KK mode expansions (it is sufficient to restrict to integers $n\ge 0$)
\begin{align}
Z^A  & = \sum_n \hat Z^A_n \cos\left(\frac{nx^2}{R}\right)+\sum'_n\tilde Z^A_n \sin\left(\frac{nx^2}{R}\right)\nonumber\\
\psi_{A +}& = \sum_n \hat \psi_{A  n+} \cos\left(\frac{nx^2}{R}\right)+\sum'_n\tilde \psi_{An +} \sin\left(\frac{nx^2}{R}\right)\nonumber\\
\psi_{A -} & = {\sum_n}  \tilde \psi_{A n-}\cos\left(\frac{nx^2}{R}\right)+\sum'_n \hat \psi_{An-} \sin\left(\frac{nx^2}{R}\right)\nonumber\\
A_\mu^{L}  & =  \sum_n A_{\mu  n} \cos\left(\frac{nx^2}{R}\right)+\sum'_n B_{\mu  n}  \sin\left(\frac{nx^2}{R}\right)\nonumber\\
A_2^{L}  & =  \sum_n A_{2 n} \cos\left(\frac{nx^2}{R}\right)+\sum'_n B_{2 n}  \sin\left(\frac{nx^2}{R}\right)\nonumber\\
A_\mu^{R}  & =  -\sum_n A_{\mu n}^t \cos\left(\frac{nx^2}{R}\right)+ \sum'_nB_{\mu n}^t \sin\left(\frac{nx^2}{R}\right)\nonumber\\
A_2^R  & =    \sum_n A_{\mu n}^t \cos\left(\frac{nx^2}{R}\right)- \sum'_nB_{\mu n}^t \sin\left(\frac{nx^2}{R}\right)\ .
\end{align}
Here $\psi_{A\pm  } = \frac12(1\pm\gamma_2)\psi_{A  }$ and a prime on the sum indicates that the $n=0$ contribution has been omitted.\footnote{ We have assumed periodic boundary conditions for the fermions, with $n\in \mathbb Z$. We could also consider supersymmetry breaking boundary conditions for the fermions by replacing $n\to r\in {\mathbb Z}+\frac12$ in the mode expansions for $\psi_{A\pm}$. However we are primarily interested in the massless modes here.} In addition a field with a hat is symmetric in its Lie-algebra indices whereas a field whereas a field with a tilde is anti-symmetric.

Although rather cumbersome one could substitute these expansions in to the  supersymmetry transformations to obtain transformation rule on the various KK modes. 
We could also substitute this ansatz into the action leading to an expression of the form
\begin{align}
S_{orb}  \sim  - & \sum_n{\rm tr}\int d^2x  \Big\{D_\mu \hat Z^A_nD^\mu \hat Z_{An}
+ \frac{n^2}{R^2}\hat Z_n^A\hat Z_{An}+D_\mu \tilde Z^A_nD^\mu \tilde Z_{An}  + \frac{n^2}{R^2}\tilde Z_n^A\tilde Z_{An} \nonumber\\
&+i  \bar{\hat\psi}^A_{n+}\gamma^\mu D_\mu \hat\psi_{An+}+ i  \bar{\hat\psi}^A_{n-}\gamma^\mu D_\mu \hat\psi_{An-}+\frac{in}{R} \bar{\hat \psi}^A_{n-} {\hat\psi}_{An+}-\frac{in}{R} \bar{\hat \psi}^A_{n+} {\hat\psi}_{An-}\nonumber\\
&+i \bar{\tilde\psi}^A_{n+}\gamma^\mu D_\mu \tilde \psi_{An+} +i \bar{\tilde\psi}^A_{n-}\gamma^\mu D_\mu \tilde \psi_{An-} 
- \frac{in}{R}\bar{\tilde\psi}^A_{n+} \tilde \psi_{An-}+ \frac{in}{R}\bar{\tilde\psi}^A_{n-} \tilde \psi_{An+}+ \ldots    \Big\}\ ,
\end{align}
involving the infinite towers of KK modes. Again it is not particularly instructive to obtain a more explicit expression for this action. However it is worth observing  that for $n\ne 0$, $(\hat\psi_{An+},\hat\psi_{An-})$ and $(\tilde \psi_{An+},\tilde \psi_{An-})$ pair up into non-chiral  fermions with masses $n/R$. However $\hat\psi_{A0+}$  and $\tilde\psi_{A-0}$ remain massless chiral fermions. 

As in any KK reduction of a gauge theory there is an infinite tower of gauge symmetries. However we will mainly be interested in the two-dimensional $U(N)$ gauge transformation that are constant along $x^2$, whose gauge field is $A_{\mu 0}$. One then sees that the fields simply transform  under  this $U(N)$ as, for example, 
\begin{equation}
Z^A\to gZ^Ag^t\ .
\end{equation}
This is a reducible representation  which can be decomposed into the symmetric and anti-symmetric representations. Thus with regards to the zero-mode gauge group fields with a hat transform in the symmetric of $U(N)$ whereas fields with a tilde transform in the anti-symmetric of $U(N)$.

Although we have put the three-dimensional theory on an orbifold, which can alternatively be thought of as a compactification on a line segment, there are no spurious boundary terms to consider that might break supersymmetry as the fields which arise from three-dimensions are all smooth at the fixed points. 
Thus the resulting action $S_{orb}$ will have $(0,6)$ supersymmetry generated by $\epsilon^{AB}_-$.

\section{An Anomaly and Its Cancellation}

 The two-dimensional action we have constructed has a  $U(N)$ gauge symmetry and a massless gauge field $A_{\mu 0}$. In addition it has massless  chiral fermions $\hat\psi_{A0+}$ in the symmetric representation of $U(N)$ and  $\tilde \psi_{A0-}$ in the anti-symmetric representation of $U(N)$. Therefore there is a gauge anomaly. 
 In particular by standard arguments there will be an anomalous variation of fermionic measure in the path integral of the form:
\begin{align}
 \delta_\omega {\rm ln}  W  &=   {\rm tr}_{\hat\psi_{A0+}}\left( \frac{1}{2\pi}\int    \omega  F_0\right)-  {\rm tr}_{\tilde\psi_{A0-}}\left( \frac{1}{2\pi}\int    \omega  F_{ 0}\right) .
&\end{align}
To continue, since $u(N)$ is not simple, we need to split $u(N)=u(1)\oplus su(N)$ and treat the $u(1)$ factor separately from $su(N)$. Let us introduce generators $t_r$, $r = 1,..,N^2-1$, for $su(N)$ and $t_0$ for $u(1)$.

For $su(N)$, since $\psi_{A0+}$ and  are in the symmetric and $\psi_{A0-}$ are in the anti-symmetric we find:
\begin{align}
 \delta_{\omega\in su(N)} {\rm ln}  W  &=  4\left(I(sym)- I(anti-sym)\right)\sum_{r\ne 0}\left(\frac{1}{2\pi} \int  \   \omega^r  F^r_{\mu\rho 0}\right)\ ,
 \end{align}
where the factor of 4 comes from the sum over the R-symmetry label   $A$ and the index  $I(R)$ is defined by the relation
\begin{equation}
{\rm tr}_R(t_rt_s) = I(R)\delta_{rs}\ .
\end{equation}
For $su(N)$ one finds
\begin{align}
I(sym) & = (N+2)I(fund)\nonumber \\
I(anti-sym) & = (N-2)I(fund)\ ,
\end{align}
where $I(fund)$ is the index in the fundamental representation of $su(N)$. Hence  
\begin{align}
 \delta_{\omega\in su(N)} {\rm ln}  W  &=  16I(fund)\sum_{r\ne 0}\left(\frac{1}{2\pi} \int  \   \omega^r  F^r_{\mu\rho 0}\right)\ .
 \end{align}

For the $u(1)$ part we first note that the normalisation of $t_0$ required to embed $U(1)$ into $U(N)$ is  
\begin{equation}
t_0^{fund}  = \frac{1}{N}1_{N\times N}\ ,
\end{equation} 
reflecting the fact that the minimal $U(1)$ charge is $1/N$. Thus $I(fund)=1/N$. 
Next we notice that  states in $\hat\psi_{A0+}$ and $\tilde\psi_{A0-}$ have twice the $U(1)$-charge of the fundamental representation, {\it i.e.} $2/N$. 
However since there are $4\times \frac12N(N+1)$ states in $\hat \psi_{A0+}$ and $4\times \frac12N(N-1)$ states in  $\tilde \psi_{A0-}$ the anomaly  is  
$4\times N\times 4/N^2 = 16 I(fund)$.
Thus the total anomaly is
\begin{align}
 \delta_\omega {\rm ln}  W  &=16  I(fund)\sum_{r}\left(\frac{1}{2\pi}\int     \omega^r  F^r_{\mu\rho 0}\right)\ .\end{align}

Since M-theory is a consistent quantum theory the orbifold must provide additional states at the fixed-point locus which cancel this anomaly. Clearly the minimal answer is that there must be 16  negative chirality fermions, corresponding to 8 at each fixed point, that are in the fundamental of $U(N)$. Thus in addition to $S_{orb}$ we must include
\begin{align}
S_{f.p.} &=  - \int_{x^2=0} d^2x\  i\bar \lambda^a_-\gamma^\mu D_\mu \lambda_{a-} -  \int_{x^2=\pi R}d^2x\ i \bar \lambda^{a'}_-\gamma^\mu D_\mu \lambda_{a'-}\ ,
\end{align}
where $a,a'=1,...,8$ and again raising/lowering an $a$ or $a'$ index is hermitian conjugation. Here the first term is localized at $x^2=0$ and the second at $x^2=\pi R$ and 
\begin{align}
D_\mu\lambda_{a-} &= \partial_\mu\lambda_{a-} - i A^L_{\mu }(x^2=0)\lambda_{a-}\nonumber\\
D_\mu\lambda_{a'-} &= \partial_\mu\lambda_{a'-} - i A^L_{\mu }(x^2=\pi R)\lambda_{a'-}\ .
\end{align}
Note that the full $A_\mu^L$ gauge field appears here, and not just the zero-mode $A_{\mu 0}$, so that the action is local. The inclusion of $\lambda_{a-}$ and $\lambda_{a'-}$  will also cancel the gravitational anomaly as observed in \cite{Horava:1995qa}.

Finally, for  the preserved supersymmetry generated by $\epsilon^{AB}_-$ we have
\begin{align}
\delta_{\epsilon}A_0^{L} = \delta_{\epsilon}A_1^{L} \ ,
\end{align}
 and therefore $S_{f.p.}$, which only involves the combination $\nabla_0-\nabla_1$, will be invariant    if we simply take 
\begin{align}
\delta_{\epsilon}\lambda_{a-}=\delta_{\epsilon}\lambda_{a'-}=0\ .
\end{align} 

Thus we propose that the full action for $N$ M2-branes on $S^1/{\mathbb Z}_2$ consists of the ABJM theory with fields restricted as in (\ref{oaction}) but also with $8$  fermions $\lambda_{a-}$ in the  fundamental of $U(N)$   localised at    the fixed point   $x^2=0$ and $8$  fermions $\lambda_{a'-}$ in the  fundamental of $U(N)$   localised at the fixed point $x^2=\pi R$. The total action is then given by
\begin{align}
S = S_{orb} + S_{f.p.}\ ,
\end{align}
and is invariant under $(0,6)$ supersymmetries generated by $\epsilon_-^{AB}$, has a  $SU(4)\times U(1)$ R-symmetry and   $SO(16)\times SO(16)$ global symmetry. The former descends from the M2-brane lagrangian whereas the  latter arises from the flavour symmetry of $8$ complex chiral fermions at each of the fixed points.

Lastly we can also consider what would happen if we took the parity operation defined in (\ref{parity1}). In this case the massless modes of the three-dimensional theory on the orbifold consist of hermitian $Z^A_0$,  $\psi_{A0+}$ and anti-hermitian $\psi_{A0-}$, {\it i.e.} $i\psi_{A0-}$ is hermitian. Furthermore the orbifold identifies  $A_{\mu 0}^L = A_{\mu 0}^R$ and $g_R(x^2)=g_L(-x^2)$ so that the zero-mode gauge group is $U(N)$ with all fields in the adjoint representation. 
Therefore there is no anomaly. The preserved supersymmetries are again  $\epsilon^{AB}_-$. Hence we cannot introduce any localized modes at the fixed points that are  in a non-trivial representation of $U(N)$ without either introducing anomalies or breaking supersymmetry. Therefore we conclude that there are no additional localized modes.

\section{The IR Theory and  $E_8\times E_8$ Heterotic Strings}

In the previous sections we constructed the orbifold theory of $N$ M2-branes on $S^1/{\mathbb Z}_2$ which involved some massless fields along with their KK towers and some additional chiral fermions that are localized to the fixed points. Let us now consider the low energy effective theory valid below the KK scale. Therefore we simply set all the non-zero KK modes to zero.
In the case of an $S^1$ compactifcation this was done in  \cite{Nastase:2010ft}  (see also \cite{Santos:2008ue,Franche:2008hr}). The result in our case is (we will return to the fermions later):
\begin{align}
S = -&{\rm tr}\int d^2x    D_\mu  \hat Y^A  D^\mu  \hat Y_A  +\frac{2}{3k^2R^2}\Upsilon^{AB}_C(\hat Y) \Upsilon_{AB}^C(\hat Y)  \nonumber \\
&+ kR\varepsilon^{\mu\nu} (A_{20} F_{\mu\nu}+A_{20}^t F^t_{\mu\nu}) + 2\pi R (A_{20} \hat Y^A-\hat Y^A A_{20}^t ) (A_{20}^t  \hat Y_A - \hat Y_A A_{20} )\nonumber \\ & +fermions\ ,
\end{align}
where
\begin{equation}
D_\mu \hat Y^A = \partial_\mu\hat Y^A - i A_{\mu 0} Y^A - i\hat Y^A A_{\mu 0}^t\ .
\end{equation}
and
\begin{equation}
\hat Y^A = \sqrt{2\pi R}\hat Z^A_0 \ ,
\end{equation}
have been rescaled to have canonical dimensions. Note that the potential term (at least for small values of $k$) has the same order as the KK masses: 
\begin{equation}
V = \frac{2}{3k^2R^2} \Upsilon^{AB}_C(\hat Y) \Upsilon_{AB}^C (\hat Y) \ .
\end{equation}
Thus at low energy, below the KK scale,   we must also restrict to the vacuum moduli space $\Upsilon^{AB}_C=0 $, which, at generic points, consists of commuting scalars.\footnote{Alternatively one could consider a large $k$ limit that would introduce a scale $1/kR<<1/R$ that is parametrically lower than the KK scale.} 

Thus the low energy effective theory below the KK scale is just the effective theory on the moduli space of vacua. Let us parameterise the moduli space as
\begin{align}\label{vac}
\hat Y^A_{vac}={\rm diag}(y^A_1,...,y^A_N)
\end{align}
This is already consistent with the orbifold action which requires a constant $Z^A$ to be symmetric. Note that had we taken the alternative parity defined in (\ref{parity1}) then  we would also require   the eigenvalues $y^A_i$ to be real. Thus the motion would be restricted to a four-dimensional hyperplane in the transverse space, corresponding the fact that the oribfold fixed point is six-dimensional, not ten-dimensional as is the case considered here. 

However we also have to mod-out by gauge transformations. For constant gauge transformations $g_R=g_L^*$ and $\hat Y^A_{vac}$ is in the symmetric representation. As with D-branes one  finds constant discrete gauge transformations that act to permute the eigenvalues. There are also continuous gauge transformations that preserve the vacuum. These are of the form $g = {\rm diag}(e^{i\theta_1},...,e^{i\theta_N})$. To examine their effect we evaluate the action we expand
\begin{align}
A^L_{\mu 0} &=  {\rm diag}(a_{\mu 1},...,a_{\mu N})\nonumber\\
A^L_{2 0} &=  {\rm diag}(a_{2 1},...,a_{2 N})\nonumber\\
A^R_{\mu 0} &=  -{\rm diag}(a_{\mu 1},...,a_{\mu N})\nonumber\\
A^R_{2 0} &=  {\rm diag}(a_{2 1},...,a_{2 N})\ .
\end{align}
We can also include the fermions by expanding them as
\begin{align} 
\hat\psi_{A0+} &= \sqrt{2\pi R}\ {\rm diag}(\chi_{A1+},...,\chi_{AN+})\nonumber\\ 
\lambda_{a-} & = \left(\begin{matrix} \lambda_{a1-} \\ \vdots\\ \lambda_{aN-}\\\end{matrix}\right)\ ,\qquad
 \lambda_{a'-} = \left(\begin{matrix} \lambda_{a'1-} \\ \vdots\\ \lambda_{a'N-}\\\end{matrix}\right)\ .
\end{align}
The action then becomes 
\begin{align}
S_{vac} = -&\sum_{i=1}^N\int d^2 x D_\mu  y_i^A D^\mu  y_{Ai}   + 2kR\varepsilon^{\mu\nu}a_{2i} F_{\mu\nu i} \nonumber\\
&+ i\bar\chi^A_{i+}\gamma^\mu D_\mu \chi_{Ai+} + i\bar \lambda^a_{i-}\gamma^\mu D_\mu \lambda_{ai-} + i\bar \lambda^{a'}_{i-}\gamma^\mu D_\mu \lambda_{a'i-}\ ,
\end{align}
where $F_{\mu\nu i}  = \partial_\mu a_{\nu  i}  -\partial_\nu  a_{\mu i} $ and the covariant derivative acts as
\begin{align}
D_\mu y^A_i &= \partial_\mu  y^A_i - 2i a_{\mu  i}y^A_i\nonumber\\
D_\mu \chi_{Ai+} &= \partial_\mu  \chi_{Ai+} - 2i a_{\mu  i}\chi_{Ai+} \nonumber \\
D_\mu \lambda_{ai-} &= \partial_\mu   \lambda_{ai-} - i a_{\mu  i} \lambda_{ai-}\nonumber\\
D_\mu \lambda_{a'i-} &= \partial_\mu  \lambda_{a'i-} - i a_{\mu  i} \lambda_{a'i-}\ .
\end{align}
The $a_{2i}$ fields can be integrated out and impose the constraint $F_{\mu\nu i}=0$. Thus we can write $a_{\mu i}=\partial_{\mu }\sigma_i$. By performing a gauge transformation we can simply set $\sigma_i=0$. Note that $\sigma_i$ is periodic with period $2\pi$ but unlike the case of uncompactified M2-branes this does not lead to a ${\mathbb Z}_k$ identification of the coordinates.
Hence the final form for the action is 
\begin{equation}
S_{vac} = \sum_{i=1}^N S_i \ ,
\end{equation}
where
\begin{align}
S_{i} = -& \int d^2 x \partial_\mu  y_i^A \partial^\mu  y_{Ai}    + i\bar\chi^A_{i+}\gamma^\mu \partial_\mu \chi_{Ai+} + i\bar \lambda^a_{i-}\gamma^\mu \partial_\mu \lambda_{ai-} + i\bar \lambda^{a'}_{i-}\gamma^\mu \partial_\mu \lambda_{a'i-}\ ,
\end{align}
is the action for a single heterotic string  consisting of 4 complex scalars $y^A$, 4 complex right-moving fermions $\chi_{A+}$ and 16 complex left-moving fermions $\lambda^a_{-}$, $\lambda^{a'}_-$. Note that due to the gauge symmetry which permutes the $i$ index the resulting effective theory is a symmetric product of $N$ free heterotic strings in ${\mathbb R}^{10}$.  

Lastly it remains to see that this is the $E_8\times E_8$ heterotic string. The distinction between the $E_8\times E_8$ and $Spin(32)/{\mathbb Z}_2$ heterotic strings arises from the choice of GSO projection \cite{Gross:1985fr,Gross:1985rr}. Here the worldsheet theory arises as the IR limit of an M2-brane theory with $SO(16)\times SO(16)$ flavour symmetry and therefore one need only impose and $SO(16)\times SO(16)$ invariant GSO projection. Indeed the $\lambda_{a-}$ and $\lambda_{a'-}$ fermions are not localised at the same orbifold fixed points so an $SO(32)$ invariant GSO would have a non-local action.   Quantization of the left-moving fermions then leads to a spacetime $E_8\times E_8$ gauge symmetry. Indeed one $E_8$ factor appears on each orbifold fixed point.

\section{Conclusions}

In this letter we have constructed an orbifold of the worldvolume theory of $N$ M2-branes on $S^1/{\mathbb Z}_2$. We showed that there was a gauge anomaly but this could be cancelled by assuming that there are 8 chiral fermions in the fundamental of $U(N)$ which are localized to each of the fixed-point planes. Taking the low energy limit of the resulting action leads to $N$ free $E_8\times E_8$ heterotic strings. The paper therefore provides a non-abelian worldvolume analogue of the classic Ho$\check{{\rm r}}$ava-Witten construction of heterotic strings \cite{Horava:1995qa,Horava:1996ma}. 

We hope that our analysis provides some insight to the M-theory origin of the  $E_8\times E_8$ spacetime gauge structure. In addition it would be interesting to see if the non-abelian theory here produces a larger spacetime symmetry gauge algebra for the case of multiple heterotic M2-branes. It would also be interesting to relate  the model we have constructed   to the $(2+1)$-dimensional matrix model description of \cite{Rey:1997hj,Kabat:1997za}.

\section*{Acknowledgements} 

I would like to thank L. Alvarez-Gaume, K. Lee and  E. Witten for helpful discussions. 
The work was supported in part by STFC grant ST/J002798/1. 
 
\bibliographystyle{utphys}
\bibliography{HetM2}

\providecommand{\href}[2]{#2}\begingroup\raggedright\begin{thebibliography}{10}

\bibitem{Horava:1995qa}
P.~Horava and E.~Witten, ``{Heterotic and type I string dynamics from
  eleven-dimensions},''
  \href{http://dx.doi.org/10.1016/0550-3213(95)00621-4}{{\em Nucl. Phys.}
  {\bfseries B460} (1996) 506--524},
\href{http://arxiv.org/abs/hep-th/9510209}{{\ttfamily arXiv:hep-th/9510209
  [hep-th]}}.

\bibitem{Horava:1996ma}
P.~Horava and E.~Witten, ``{Eleven-dimensional supergravity on a manifold with
  boundary},'' \href{http://dx.doi.org/10.1016/0550-3213(96)00308-2}{{\em Nucl.
  Phys.} {\bfseries B475} (1996) 94--114},
\href{http://arxiv.org/abs/hep-th/9603142}{{\ttfamily arXiv:hep-th/9603142
  [hep-th]}}.

\bibitem{Rey:1997hj}
S.-J. Rey, ``{Heterotic M(atrix) strings and their interactions},''
  \href{http://dx.doi.org/10.1016/S0550-3213(97)00428-8}{{\em Nucl. Phys.}
  {\bfseries B502} (1997) 170--190},
\href{http://arxiv.org/abs/hep-th/9704158}{{\ttfamily arXiv:hep-th/9704158
  [hep-th]}}.

\bibitem{Kabat:1997za}
D.~N. Kabat and S.-J. Rey, ``{Wilson lines and T duality in heterotic M(atrix)
  theory},'' \href{http://dx.doi.org/10.1016/S0550-3213(97)00605-6}{{\em Nucl.
  Phys.} {\bfseries B508} (1997) 535--568},
\href{http://arxiv.org/abs/hep-th/9707099}{{\ttfamily arXiv:hep-th/9707099
  [hep-th]}}.

\bibitem{Gross:1985fr}
D.~J. Gross, J.~A. Harvey, E.~J. Martinec, and R.~Rohm, ``{Heterotic String
  Theory. 1. The Free Heterotic String},''
\href{http://dx.doi.org/10.1016/0550-3213(85)90394-3}{{\em Nucl. Phys.}
  {\bfseries B256} (1985) 253}.

\bibitem{Gross:1985rr}
D.~J. Gross, J.~A. Harvey, E.~J. Martinec, and R.~Rohm, ``{Heterotic String
  Theory. 2. The Interacting Heterotic String},''
\href{http://dx.doi.org/10.1016/0550-3213(86)90146-X}{{\em Nucl. Phys.}
  {\bfseries B267} (1986) 75}.

\bibitem{Berman:2009xd}
D.~S. Berman, M.~J. Perry, E.~Sezgin, and D.~C. Thompson, ``{Boundary
  Conditions for Interacting Membranes},''
  \href{http://dx.doi.org/10.1007/JHEP04(2010)025}{{\em JHEP} {\bfseries 04}
  (2010) 025},
\href{http://arxiv.org/abs/0912.3504}{{\ttfamily arXiv:0912.3504 [hep-th]}}.

\bibitem{Chu:2009ms}
C.-S. Chu and D.~J. Smith, ``{Multiple Self-Dual Strings on M5-Branes},''
  \href{http://dx.doi.org/10.1007/JHEP01(2010)001}{{\em JHEP} {\bfseries 01}
  (2010) 001},
\href{http://arxiv.org/abs/0909.2333}{{\ttfamily arXiv:0909.2333 [hep-th]}}.

\bibitem{Kim:2014dza}
J.~Kim, S.~Kim, K.~Lee, J.~Park, and C.~Vafa, ``{Elliptic Genus of
  E-strings},''
\href{http://arxiv.org/abs/1411.2324}{{\ttfamily arXiv:1411.2324 [hep-th]}}.

\bibitem{Haghighat:2014pva}
B.~Haghighat, G.~Lockhart, and C.~Vafa, ``{Fusing E-strings to heterotic
  strings: E+E→H},'' \href{http://dx.doi.org/10.1103/PhysRevD.90.126012}{{\em
  Phys. Rev.} {\bfseries D90} no.~12, (2014) 126012},
\href{http://arxiv.org/abs/1406.0850}{{\ttfamily arXiv:1406.0850 [hep-th]}}.

\bibitem{witten:strings}
E.~Witten, ``{An Overview of Worldsheet Brane Anomalies },'' {\em {Talk given
  at Strings 2015}} .
  \url{{https://strings2015.icts.res.in/talkDocuments/6-2.00-2.30-Edward-Witten.pdf}}.

\bibitem{Aharony:2008ug}
O.~Aharony, O.~Bergman, D.~L. Jafferis, and J.~Maldacena, ``{N=6 superconformal
  Chern-Simons-matter theories, M2-branes and their gravity duals},''
  \href{http://dx.doi.org/10.1088/1126-6708/2008/10/091}{{\em JHEP} {\bfseries
  10} (2008) 091},
\href{http://arxiv.org/abs/0806.1218}{{\ttfamily arXiv:0806.1218 [hep-th]}}.

\bibitem{Bagger:2008se}
J.~Bagger and N.~Lambert, ``{Three-Algebras and N=6 Chern-Simons Gauge
  Theories},'' \href{http://dx.doi.org/10.1103/PhysRevD.79.025002}{{\em Phys.
  Rev.} {\bfseries D79} (2009) 025002},
\href{http://arxiv.org/abs/0807.0163}{{\ttfamily arXiv:0807.0163 [hep-th]}}.

\bibitem{Gustavsson:2007vu}
A.~Gustavsson, ``{Algebraic structures on parallel M2-branes},''
  \href{http://dx.doi.org/10.1016/j.nuclphysb.2008.11.014}{{\em Nucl. Phys.}
  {\bfseries B811} (2009) 66--76},
\href{http://arxiv.org/abs/0709.1260}{{\ttfamily arXiv:0709.1260 [hep-th]}}.

\bibitem{Bagger:2007jr}
J.~Bagger and N.~Lambert, ``{Gauge symmetry and supersymmetry of multiple
  M2-branes},'' \href{http://dx.doi.org/10.1103/PhysRevD.77.065008}{{\em Phys.
  Rev.} {\bfseries D77} (2008) 065008},
\href{http://arxiv.org/abs/0711.0955}{{\ttfamily arXiv:0711.0955 [hep-th]}}.

\bibitem{Bagger:2012jb}
J.~Bagger, N.~Lambert, S.~Mukhi, and C.~Papageorgakis, ``{Multiple Membranes in
  M-theory},'' \href{http://dx.doi.org/10.1016/j.physrep.2013.01.006}{{\em
  Phys. Rept.} {\bfseries 527} (2013) 1--100},
\href{http://arxiv.org/abs/1203.3546}{{\ttfamily arXiv:1203.3546 [hep-th]}}.

\bibitem{Nastase:2010ft}
H.~Nastase and C.~Papageorgakis, ``{Dimensional reduction of the ABJM model},''
  \href{http://dx.doi.org/10.1007/JHEP03(2011)094}{{\em JHEP} {\bfseries 03}
  (2011) 094},
\href{http://arxiv.org/abs/1010.3808}{{\ttfamily arXiv:1010.3808 [hep-th]}}.

\bibitem{Santos:2008ue}
M.~A. Santos and I.~V. Vancea, ``{New two-dimensional massless field theory
  from Bagger-Lambert-Gustavsson model},''
  \href{http://dx.doi.org/10.1142/S0217732309030746}{{\em Mod. Phys. Lett.}
  {\bfseries A24} (2009) 2275--2284},
\href{http://arxiv.org/abs/0809.0256}{{\ttfamily arXiv:0809.0256 [hep-th]}}.

\bibitem{Franche:2008hr}
P.~Franche, ``{Reduction of the N=8 BLG and N=6 BL Theories to 2D Effective
  Field Theories},''
\href{http://arxiv.org/abs/0811.1443}{{\ttfamily arXiv:0811.1443 [hep-th]}}.

\end{thebibliography}\endgroup

\end{document}